# A Trust Management System for the IoT domain

Christos-Minas Mathas, Costas Vassilakis, Nicholas Kolokotronis
Department of Informatics and Telecommunications
University of the Peloponnese
Greece
mathas.ch.m@uop.gr, costas@uop.gr, nkolok@uop.gr

*Abstract*—In modern internet-scale computing, interaction between a large number of parties that are not known a-priori is predominant, with each party functioning both as a provider and consumer of services and information. In such an environment, traditional access control mechanisms face considerable limitations, since granting appropriate authorizations to each distinct party is infeasible both due to the high number of grantees and the dynamic nature of interactions. Trust management has emerged as a solution to this issue, offering aids towards the automated verification of actions against security policies. In this paper, we present a trust- and risk-based approach to security, which considers status, behavior and associated risk aspects in the trust computation process, while additionally it captures user-to-user trust relationships which are propagated to the device level, through user-to-device ownership links.

## I. INTRODUCTION

In modern internet-scale computing, interaction between a large number of parties that are not known a-priori is predominant, with each party functioning both as a provider and consumer of services and information. In such an environment, traditional access control mechanisms face considerable limitations, since granting appropriate authorizations to each distinct party is infeasible both due to the high number of grantees and the dynamic nature of interactions. The concept of trust management has emerged as a solution to this issue. [1] defines trust management as an aid to the automated verification of actions against security policies. According to this definition, an action is allowed if the credentials presented are deemed sufficient, without the need to state or verify the actual identity of the interacting party; in this respect, symbolic representation of trust is separated from the actual person (or the person's digital agent). Later research has replaced the examination of credentials (which could be considered as *pseudonymized identities*, limiting hence the benefits of introducing trust management [2]) to the examination of a *set of properties*, which can be proven by an interacting party through the presentation of a set of digital certificates [3]. Under this scheme, the original set of trust management system elements identified in [4] is modified to include (a) *security policies*, which are a set of trust assertions that are considered "ground truth" and are trusted unconditionally; (b) *trust-related properties*, which represent aspects of interacting parties that are relevant to the application of security policies, typically, examined as antecedents of rules that comprise a security policy; and (c) *trust relationships*, which are a special kind security policy.

The computation of the trust level for an interaction peer may involve all observable aspects for this peer: this spans across (a) the *security aspects of the interaction peer*, including the current assessment of the peer's integrity status (known tampering of firmware, operating system, critical files; patching level; etc.) and (b) *behavioral aspects of the interaction peer*, mainly focused on whether the interaction peer (i) operates according to its predefined usage description and (ii) deviates from its normal behavior.

Services, information and resources that need to be protected through trust management or other relevant approaches are ultimately *assets*, and each asset has a *value* for its owner. Furthermore, assets are exposed to a number of threats; each such threat constitutes a risk for demotion of the respective assets' value. To this end, effective asset protection entails the assessment of the risk posed by each interaction and deciding on the defensive actions that possibly need to be taken on the grounds of this assessment.

Trust and risk assessment are closely linked, since analysis of information security risks entails the assessment of the realistic likelihood that the identified risks will occur [6], and this probability in turn depends on the level of trust placed on systems that could potentially be threat agents: this is reflected on the definition of trust listed in [7] according to which "Trust is the willingness of a party to be vulnerable to the action of another party, based on the expectation that the other will perform a particular action important to the trustor, irrespective to the ability to monitor or control that other party". Taking this into account, we conclude that trust moderates the level of risk, through the belief that a trusted system is not bound to effectively function as a threat agent. Under this viewpoint, the trust assessment for a system is a critical parameter to be taken into account when performing risk assessment.

Finally, contemporary attack methods entail complex multi-stage, multi-host attack paths, where each path represents a chain of exploits used by an attacker to break into a network [8]. Attack graphs enable the comprehensive risk analysis within a network, considering cause-consequence relationships between different network states; furthermore, the likelihood that such relationships would be exploited can be also taken into account [9].

In this paper, we present a trust- and risk-based approach to security, within which trust establishment and risk assessment encompass all the above listed aspects, providing thus a holistic trust management and risk assessment view. The proposed approach advances the state-of-the-art in trust models by extending the trust computation procedure to include status and associated risk aspects, beyond the behavior aspect which is typically used in trust management systems. Additionally, the proposed approach captures user-to-user trust relationships, which are propagated to the device level through user-to-device ownership links.

The rest of this paper is structured as follows: section II overviews related work, while section III presents the trust management model. Section IV presents the trust management system operational context, while section V concludes the paper and outlines future work.

## II. Related Work

This section overviews the trust models that have been proposed by the literature trying to find an effective and efficient trust computation method. In service-oriented networks, an IoT device acting as a service requester needs a way of evaluating which of its peers can be trusted to provide it with the requested service, while taking into consideration the energy demands of carrying out such evaluation. This is the challenge that trust management models are aiming to solve. For each model, the approaches adopted for trust composition (QoS, Social) and trust computation are presented, while salient features of the models are summarized. Trust composition refers to the components the given model takes into consideration. The components include Quality of Service (QoS) and Social trust. QoS trust refers to the evaluation of a node based on its capability to deliver the requested service. Social trust refers to the social relationship between owners of IoT devices.

The model proposed in [10] considers Community of interest (CoI) based social IoT (SIoT) systems. Devices have owners, each owner has many devices and keeps a friends list. Nodes belonging to similar communities are more likely to have similar interests or capabilities. Both QoS and Social trust composition are considered, including three trust properties: honesty (QoS), cooperativeness (QoS) and community-interest (social). Trust values can occur from direct observations and when such observations are not available, from recommendations. [10] follows a distributed scheme, while for trust aggregation the weighted sum technique is used.

The work in [11] is very similar to [10]. The main difference is that it follows a scenario-based trust computation scheme rather than a general approach for the computation of overall trust. The model is validated against in two real-world scenarios, namely, "Smart City Air Pollution Detection" and "Augmented Map Travel Assistance". The work in [12] also resembles the model in [10], employing however Bayesian inference for direct trust, while weighted sums are used to aggregate recommendations into indirect trust. Additionally, [12] introduces an efficient storage management strategy suitable for large-scale IoT systems. [13] extends the model in [12] by (a) considering friendship and social contact in the evaluation of recommenders and (b) combining direct with indirect trust (recommendations) to form the overall trust.

The model in [14] considers the service satisfaction at a given time from a specific service provided by a node (a QoS property). Trust is defined as: the Direct Trust value, the Recommended Trust value if the node calculating the trust value had no interaction with the rated service/node and thus the Direct Trust value can't be calculated, or as a predefined Ignorance Value if the rated node is joining the cloud environment for the first time. The weights assigned for each Direct Trust value are calculated using the number of positive interactions between the node calculating the trust value and the node whose rating is considered in the weight calculation, and the Satisfaction Level –a factor dependent on availability, recovery time, connectivity and peak-load performance as defined in the service agreement.

The model proposed in [15] is a distributed version of the model proposed by [16]. Both models consider ratings given to a specific node and service. [15] operate in three phases: 1) every node announces its presence to its neighbors while also keeping a record of its neighbors, 2) a node requests a service from a neighboring node and rates the response as positive or negative, and 3) the node calculates and stores the trust value of its neighbor.

The work in [17] proposes two models, labeled as the subjective model and the objective model. The two proposed models, consider seven parameters: service ratings, number of transactions per node –to detect nodes with an abnormal number of transactions, node credibility, transaction factor –separating important transactions to avoid trust to be built solely by many small transactions, computation capacity –as "smarter" nodes can be better suited to become malicious, relationship factor –the type of relation between two nodes, and finally the notion of centrality –as a node with many connections or involved in many transactions takes a central role in the network.

## III. The Trust Management System

In this section we present the proposed trust management system (TMS) for the IoT domain. Firstly, in subsection III.A the entities considered in the TMS design and the relationships among them are described, while subsection III.B describes the methods used to compute the trust level for the IoT devices.

### A. Entities considered in the TMS, and their relationships

Fig. 1 depicts the entities considered by the proposed trust model in the Smart home/SOHO/Enterprise network IoT environments and the relationships between these elements. The elements are as follows:

*Devices*, that operate within the considered environment. A device may be either a stationary device that is bound to a specific network or a mobile device that may participate in different networks at different time instants, in an ad-hoc basis.

*Users*, who own devices. A single user may own multiple devices. Users may develop trust relationships between them;

trust relationships between users are *directed*, *not necessarily symmetrical*, *not transitive* and *weighted*, i.e.:

- Some user $u_1$ declares to trust some other user $u_2$, providing a *trust level*, expressing $u_1$'s confidence that $u_2$ will not act in a way that is harmful for $u_1$.
- The assertion of trust towards $u_2$ made by $u_1$ does not imply in any way that $u_2$ also trusts $u_1$, expressing the fact that trust may not be reciprocated [15].
- Trust is not transitive: if $u_1$ trusts $u_2$ and $u_2$ trusts $u_3$, no assumption is made that $u_1$ trusts $u_3$.

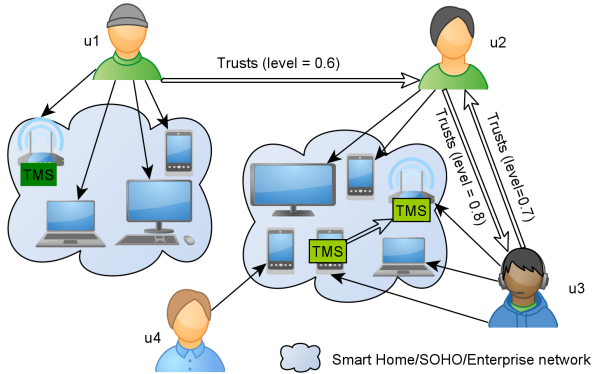

Figure 1. The entities in the IoT environment and their relationships

*Trust Management System* (TMS) *instances*: these are software agents operating within the considered environment and implement functionalities for computing trust levels towards devices. To compute the value of trust towards a device, the TMS synthesizes multiple pieces of information, either explicitly provided or gathered through observations. These pieces of information are: (a) the status of the device, (b) the behaviour of the device, (c) the risk associated with the device and (d) the trust relationship between the owners of the devices. These pieces of information are discussed in the following subsections.

*1) Device status*

The status of the device encompasses (1) information regarding the *device integrity*, i.e. the extent to which the device is known to run legitimate firmware/operating system/software under a validated configuration, as contrasted to the case that these device elements have been tampered with; and (2) information regarding the *device resilience*, i.e. the extent to which the device firmware/operating system/software/configuration are known to be free of security vulnerabilities, as contrasted to the case that such known vulnerabilities exist.

*2) Device behaviour*

The device behaviour dimension includes information regarding whether:

- the device has been detected to launch attacks, or be target of attacks.
- the device's resource usage metrics are within a pre-determined range which is considered to be "normal" or deviate from it. The metrics can pertain to any observable aspect of resource usage metrics, e.g. CPU load, network usage or disk activity. Practically, any class of system metrics that can be quantified, and for which baseline metrics can be created so as to allow computation of deviations from the baselines is eligible for incorporation within this dimension.
- the device complies with pre-specified behaviour which has been whitelisted as benign. MUD specification [19] files are the most prominent source of such information, albeit their adoption and manufacturer support is lagging behind expectations.

*3) Associated risk*

Devices within the IoT may be attacked, and some attacks may be successful. The probability that each device is finally compromised can be computed taking into account only technical information, such as the reachability of the device and the vulnerabilities present on it, and attack graphs are a prominent tool for supporting such computations [20]. However, not all compromises have the same level of impact on the organization/person owning the device: the level of impact is moderated by the perceived value of the device. The perceived value of the device in turn is moderated by (a) the assets that the device hosts (e.g. a database) or the value/criticality of the dependent processes that the device supports (e.g. a temperature sensor may support a simple temperature reading application or the automated cooling system of a nuclear reactor).

Furthermore, in the context of sophisticated, multi-staged attacks, a compromised device $d$ may be used as a stepping-stone, enabling the attackers to launch attacks against other devices which are reachable from $d$ and may be otherwise unreachable (or harder to reach), if $d$ were not compromised. Notably, the devices that are reachable from $d$ contain themselves assets that have a business value, and the technical probability that these devices are compromised in the context of multi-staged attacks can be jointly considered with the respective business values to provide an additional aspect of the risk associated with device $d$.

The associated risk dimension combines the above-mentioned aspects into a single, metric expressing the business risk associated with a device.

*4) User-to-user trust relationships*

The proposed trust model considers trust relationships between the owner of the device that performs trust assessment and the owner of the device, for which the trust evaluation is conducted. This aspect allows the propagation of the trust between users to the level of the devices they own.

The three separate trust dimensions, i.e. (i) status-based, (ii) behaviour-based and (iii) associated risk-based trust, are synthesized by the TMS instances into a single comprehensive trust assessment; this trust assessment is further moderated considering user-to-user trust relationships.

*B. Trust computation*

As described in subsection III.A, the TMS synthesizes a comprehensive trust score, taking for each device into account (a) its status, (b) its behaviour, (c) the associated risk and (d) user-to-user trust relationships. In this subsection, we describe in detail the methods used for computing the different dimensions of the device trust, and synthesizing these dimensions into a comprehensive trust score.

*1) Computation of the status-based trust score*

The trust-based score of a device *D* comprises the *integrity aspect* and the *vulnerability aspect*.

The *integrity* aspect relates to whether the software components of the device (firmware, operating systems and generic software applications) are integral or have been tampered with; this status aspect is denoted as $SBT_I(D)$. When some device has been detected to be compromised, the TMS sets $SBT_I(D)$ to zero. $SBT_I(D)$ is restored to one when the health of a device is found to be restored.

On the other hand, the *vulnerability* aspect relates to whether the software bears weaknesses which can be exploited to compromise the device. In this context, only vulnerabilities having a *network* or *adjacent* attack vector [21] (i.e. vulnerabilities that can be exploited remotely) are considered. Each vulnerability has an associated impact score, expressing the impact of the vulnerability, taking into account the effect that it may have on the value of the device as well as the exploitability of the vulnerability [21]. Therefore, the overall vulnerability impact metric for device *D*, denoted as *OVIM(D)*, can be calculated as

$$OVIM(D) = \sum_{v \in vul(D)} ne(v) * \frac{im(v)}{10} \quad (1)$$

where:
- *vul (D)* is the set of vulnerabilities present on device *D*;
- *ne(v)* is equal to 1 if vulnerability *v* is remotely exploitable or zero, otherwise; and
- *im(v)* is the impact metric for vulnerability *v*; the value of *im(v)* is divided by 10 to normalize its range into [0, 1], since the CVSS specification [21] designates a range [0, 10].

The value of *OVIM(D)* is normalized in the range [0, 1] to produce the status score related to the aspect of vulnerabilities for device *D* using equation (2):

$$SBT_V(D) = 1 - e^{-OVIM(D)} \quad (2)$$

The value of SBTV(D) is modified when new vulnerabilities are associated with the device and when vulnerabilities are mitigated (e.g. by installation of a patch, removal of the vulnerable software components etc.).

Finally, the partial status-based scores SBTI(D) and SBTV(D) are combined to formulate an overall status-based trust assessment for D, which is denoted as SBT(D). This is accomplished using equation (3).

$$SBT(D) = SBT_I(D) * SBT_V(D) \quad (3)$$

*2) Computation of the behaviour-based trust score*

The behaviour-based trust score for a device *D* comprises three distinct aspects, namely *compliance*, *normal behaviour* and *malicious activities*.

*Compliance*, corresponds to whether *D* in accordance with some rules which describe benign behaviour for the particular device; the score for this behavioural aspect is denoted as $BBT_C(D)$. This aspect mainly applies to network traffic, and in this context the MUD specification is the prevalent approach [19], defining compliance through a set of rules designating the allowed traffic flows. When *D* sends traffic that does not adhere to such rules, it is flagged as *non-compliant* and $BBT_C(D)$ is set to zero. However, the non-compliance penalty should not remain indefinitely, since the deviation may be coincidental: for instance, the system administrator could issue a command initiating a non-compliant traffic flow, in an otherwise benign system. To guard against cases of indefinite demotions of compliance-related trust scores, the TMS restores $BBT_C(D)$ at some specific rate, which is moderated through a respective system parameter, $TSRR_{compliance}$. Should the device continue to exhibit non-compliant behaviour, $BBT_C(D)$ will be again set to zero.

*Normal behaviour*, corresponds to whether the observable aspects of resource usage metrics exhibited by the device fall in the range that is typically exhibited by the device, as determined by the collection and classification of historical device behaviour data; the score for this behavioural aspect is denoted as $BBT_N(D)$. When abnormal behaviour is detected, the TMS decreases $BBT_N(D)$; the deduction made to $BBT_N(D)$ is equal to the degree of deviation from the nominal metrics: in particular, the degree of deviation is computed as

$$devDegree = \max\left(\frac{detectedMaxVal - nominalHighEnd}{detectedMaxVal}, 0\right) \quad (4)$$

where *detectedMaxVal* is the detected maximum value for the metric and *nominalHighEnd* is the high end of the nominal range of the metric.

Similarly to the case of compliance, the value of $BBT_N(D)$ is gradually restored at some specific rate, to guard against coincidental deviations (e.g. an unanticipated update operation or a backup operation producing higher bulks of data transfers than nominal volumes; or when a device is under a DDoS attack, the network metrics -and probably CPU metrics- may deviate from the respective nominal values), the TMS restores $BBT_N(D)$ at some specific rate, which is moderated through a respective system parameter ($TSRR_{nominality}$). Should the device continue to exhibit deviant behaviour, $BBT_N(D)$ will be repetitively reduced and thus maintained at low levels.

Finally, *malicious activities*, corresponds to the detection of attacks being launched from *D*; the score for this behavioural aspect is denoted as $BBT_M(D)$. When launching of attacks is detected, the TMS sets $BBT_M(D)$ to zero. Contrary to the cases of $BBT_M(D)$ and $BBT_M(D)$, the TMS *does not* restore the value of $BBT_M(D)$, since the launching of an attack is deemed improbable to be coincidental. $BBT_M(D)$ can only be restored when the health of a device is explicitly designated to be restored (typically through manual intervention).

The TMS synthesizes the values pertaining to the different aspects of the behaviour-based trust dimension into a single, comprehensive score for behaviour-based trust, which is

denoted as *BBT(D)*. The value of *BBT(D)* is computed according to formula (5):

$$BBT(D) = BBT_C(D) * BBT_N(D) * BBT_M(D) \quad (5)$$

According to formula (5), a major demotion of the score of any of the behavioural aspects leads to a low value for the behaviour-based trust dimension.

*3) Computation of the associated risk-based trust score*

The risk-based trust score dimension combines the technical probability that a machine is compromised with the level of the damage that would be sustained to the owner of the machine/infrastructure as a result of this compromise, to accommodate a business-oriented security aspect, in line with the information system risk assessment model [22].

In order to compute the associated risk-based trust score for device *D* the TMS performs steps described in the following paragraphs.

Firstly, it combines the probability that *D* is compromised with the perceived impact of the machine compromise, as explicitly entered by the user. This is accomplished using the risk assessment matrix (adapted from [23]) shown in Fig. 2.

| Severity level | Probability of occurrence | | | | |
|---|---|---|---|---|---|
| | Highly probable | Probable | Medium | Remote | Improbable |
| Catastrophic | Catastrophic | Catastrophic | Catastrophic | Serious | Medium |
| Severe | Catastrophic | Catastrophic | Serious | Medium | Low |
| Normal | Catastrophic | Serious | Medium | Low | Negligible |
| Minor | Serious | Medium | Low | Negligible | Negligible |
| Negligible | Medium | Low | Negligible | Negligible | Negligible |

Figure 2. Risk assessment matrix

The use of the table in Fig. 2 results in the computation of a fuzzy risk label constituting the *fuzzy label singular risk assessment for device* which is denoted as $SRA_{FL}(D)$. Fuzzy labels can be converted to numeric ratings by dividing the range [0,1] in a number of strata (0.0; 0.25; 0.5; 0.75; 1.0) and mapping fuzzy labels to the corresponding stratum value. This constitutes the *numerical singular risk assessment for D*, and is denoted as $SRA_L(D)$, i.e.:

$$SRA_L(D) = \begin{cases} 0 & if\ SRA_{FL}(D) = Negligible \\ 0.25 & if\ SRA_{FL}(D) = Low \\ 0.5 & if\ SRA_{FL}(D) = Medium \\ 0.75 & if\ SRA_{FL}(D) = Serious \\ 1 & if\ SRA_{FL}(D) = Catastrophic \end{cases} \quad (6)$$

Secondly, the TMS considers the fact that if *D* is compromised in a fashion that allows remote code execution, then *D* can be used by the attacker as a stepping-stone to commit attacks against other machines within the protected infrastructure, leading thus to the potential of additional impact being incurred on the organization and, consequently, higher risk levels. To accommodate this dimension, the TMS considers (a) the probability that *D* is compromised in a fashion that allows remote code execution, (b) the neighbouring devices of *D*, (c) the vulnerabilities of each of the neighbouring devices that would permit remote exploitation and the severity of each one of them and (d) the perceived value of each of the neighbouring devices. To compute this dimension, the TMS first computes the cumulative effect on the risk on neighbouring infrastructure stemming from the potential compromise as

$$CCEN(D) = \sum_{n \in neighbours(D)} SRA_L(n) \quad (7)$$

The value of *CCEN(D)* computed using equation (7) may be arbitrarily high, while additionally it does not consider the base probability that *D* is compromised. To normalize the *CCEN(D)* in the range [0, 1] and accommodate the probability that *D* is compromised, the TMS synthesises the value of *CCEN(D)* and the probability *PRC(D)* that *D* is compromised to compute the *amortized cumulative compromised effect on neighbouring infrastructure* using equation (8):

$$ACCEN(D) = PRC(D) * (1 - e^{-CCEN(D)}) \quad (8)$$

Finally, the values of $SRA_N(D)$ and *ACCEN(D)* are combined to compute the overall risk assessment for *D*:

$$ABT(D) = 1 - \max(SRA_N(D), ACCEN(D)) \quad (9)$$

*4) Synthesizing the status-based, behaviour-based and associated risk-based scores*

The three dimensions of trust, whose calculation was presented in sections III.B.1-III.B.3 are synthesized, in order to produce a comprehensive trust score, which considers all trust-related aspects of the device. This comprehensive score reflects the *local view* of the TMS computing the trust level, and will be referred to as *local trust assessment* (LTA).

Several methods for the combination of individual trust scores can be employed, including *simple additive weighting*, *fuzzy simple additive weighting* and *multiplicative* [25]. At this stage, the TMS adopts the simple additive weighting method, according to which a weight is attached to each of the dimensions, with the sum of weights being equal to 1. Effectively, for the case of the TMS, three weights $w_s$, $w_b$ and $w_a$ would need to be defined, associated with the status, behaviour and associated risk, respectively, with $0 \leq w_s$, $w_b$, $w_a \leq 1$ and $w_s + w_b + w_a = 1$. Then, the local trust assessment for device *D*, *LTA(D)* would be:

$$LTA(D) = w_s * SBT(D) + w_b * BBT(D) + w_a * ABT(D) \quad (10)$$

Considering the values of $w_s$, $w_b$ and $w_a$, we may note that the behaviour trust score is based on evidence on the activity of the device; on the other hand, the presence of vulnerabilities on a device, while undesirable, may or may not lead to its compromise (depending on a number of factors such as the reachability of the device or the perceived value of the device for attackers). Consequently, we expect that $w_b > w_s$. Similarly, the associated risk dimension pertains to events that may occur, and correspondingly $w_b > w_a$.

The use of alternative methods for the combination of individual trust scores will be investigated in our future work.

*5) Incorporating user-to-user trust relationships and computing the final trust score*

The trust assessment of a device -as computed in the previous subsection- is an objective measure, synthesizing the status, behaviour and associated risk dimensions observed by the TMS. In the final step, the TMS takes into account the trust relationships between the users, and in particular between the owner of the TMS $U$ and the owner of device $D$, who will be denoted as *Owner(D)*. The trust level between two users $U_1$ and $U_2$ is denoted as $UT(U_1, U_2)$ and is computed as follows:

$$UT(U_1, U_2) = \begin{cases} 1 & \text{if } U_1 = U_2 \\ ETS(U_1, U_2) & \text{if } U_1 \text{ has explicitly established a} \\ & \text{trust relationship towards } U_2 \text{ with} \\ & \text{a trust level equal to } ETS(U_1, U_2) \\ UT_{known} & \text{If the identity of owner(D) is known} \\ & \text{but no trust relationship from } U_1 \\ & \text{to } U_2 \text{ has been established} \\ UT_{unknown} & \text{If the identity of owner(D) is} \\ & \text{not known} \end{cases} \quad (11)$$

$UT_{known}$ and $UT_{unknown}$ are parameters of the TMS, which regulate the trust level assigned to devices for which no direct trust relationship has been established. Users with known identities are expected to be assigned higher trust levels, under the rationale that known can be held accountable for the activities of their devices, and therefore the probability that such a device is deliberately launching attacks is limited.

The user trust level computed by equation (11) is used by the TMS to moderate the trust assessment computed by equation (10) and produce the final trust score as follows:

$$TS(D) = LTA(D) * UT(Owner(T), Owner(D)) \quad (12)$$

## IV. TMS OPERATIONAL CONTEXT

Fig. 3 illustrates the operational context of the TMS. The TMS follows the security information and event management (SIEM) paradigm [27], gathering information from other components within its operational context and exploiting them to synthesize comprehensive trust assessments. The components from which SIEM information are sourced are described in the following paragraphs.

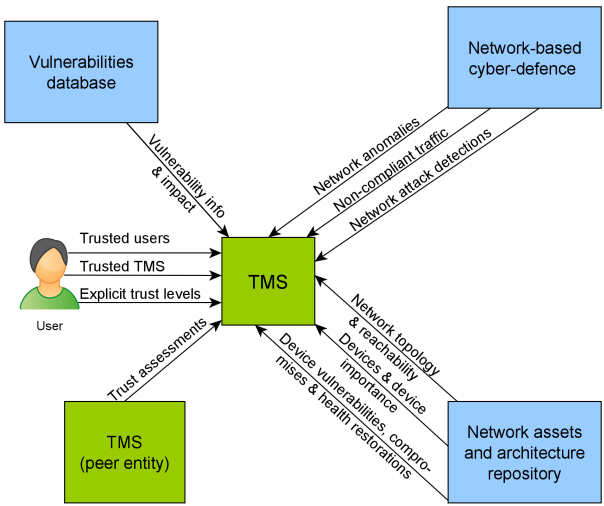

Figure 3. The operational context of the TMS

*Network-based cyber-defence*: This component analyses network traffic within the defended network perimeter, identifying attacks, network anomalies (e.g. deviations from typical traffic behaviour for a device), as well as cases when devices generate non-compliant traffic. Algorithms developed for IDS and IPS systems [28] can support the provision of this information.

*Network assets and architecture repository*: this component provides information about the network of discourse and the devices included therein. This information comprises *device importance* (i.e. the value of the assets hosted on the device, as perceived by the asset owner), the device vulnerabilities (either self-reported by an agent running on the device or remotely sensed through tools such as the ones listed in [29]) and the cases that devices are compromised or their health is restored (determined e.g. through remote attestation [30]). This component additionally provides information about the network topology and reachability information, allowing the TMS to identify cases where some device can be used as a stepping stone to launch attacks on other devices, in the context of a multi-stage attack path.

*A vulnerabilities database,* providing information about the impact of the vulnerabilities that are detected on devices; as described in section III.B, this information is used to compute the status-based and the risk-based trust dimension scores. The NIST National Vulnerability Database (NVD) [31] is a prominent data source that can be used to this effect.

## V. CONCLUSIONS

In this paper we have presented a trust management system for the IoT domain. The proposed system formulates trust assessments for devices by synthesizing three important views on devices' security, namely the behavior, status and associated risk, providing thus a holistic trust and risk assessment view. Additionally, the TMS captures user-to-user trust relationships which are propagated to the device level, through user-to-device ownership links.

A preliminary evaluation for the proposed TMS has been conducted in [32], where it has been shown to be resilient against TMS-related attacks, while it can also provide timely information on device trust changes that are of importance. Further, simulation-based evaluations are planned to be conducted within the context of our future work.

Our future work will additionally include the cooperation between TMS instances, as well as the consideration of application-level trust-related events for the computation of trust. In order to facilitate the generation and dissemination of application-level trust-related events, the development of a suitable framework will be considered. A rigorous evaluation of the TMS in the context of real-world attack scenarios, and the formalisation of the link between trust and risk assessment by means of quantitative formalisms (e.g., Bayesian Networks) are also planned.

ACKNOWLEDGMENT

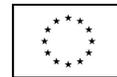 This project has received funding from the European Union's Horizon 2020 research and innovation programme under grant agreement no. 786698. The work reflects only the authors' view and the

REFERENCES

[1] M. Blaze, J. Ioannidis, and A. D. Keromytis, "Experience with the KeyNote Trust Management System: Applications and Future Directions," 2003, pp. 284–300.

[2] C. A. Ardagna, E. Damiani, S. De Capitani di Vimercati, S. Foresti, and P. Samarati, "Trust Management," in *Security, Privacy, and Trust in Modern Data Management*, Berlin, Heidelberg: Springer Berlin Heidelberg, 2007, pp. 103–117.

[3] P. A. Bonatti and P. Samarati, "A uniform framework for regulating service access and information release on the Web," *J. Comput. Secur.*, vol. 10, no. 3, pp. 241–271, Jul. 2002, doi: 10.3233/JCS-2002-10303.

[4] M. Petković and W. Jonker, Eds., *Security, Privacy, and Trust in Modern Data Management*. Berlin, Heidelberg: Springer Berlin Heidelberg, 2007.

[5] M. Srivatsa, L. Xiong, and L. Liu, "TrustGuard," in *Proceedings of the 14th international conference on World Wide Web - WWW '05*, 2005, p. 422, doi: 10.1145/1060745.1060808.

[6] Joint Technical Committee ISO/IEC JTC 1, "International standard NEN-ISO/IEC 27001: Information technology - Security techniques - Information security management systems - Requirements," 2013.

[7] R. C. Mayer, J. H. Davis, and F. D. Schoorman, "An Integrative Model of Organizational Trust," *Acad. Manag. Rev.*, vol. 20, no. 3, p. 709, Jul. 1995, doi: 10.2307/258792.

[8] X. Ou, W. F. Boyer, and M. A. McQueen, "A scalable approach to attack graph generation," in *Proceedings of the 13th ACM conference on Computer and communications security - CCS '06*, 2006, pp. 336–345, doi: 10.1145/1180405.1180446.

[9] N. Poolsappasit, R. Dewri, and I. Ray, "Dynamic Security Risk Management Using Bayesian Attack Graphs," *IEEE Trans. Dependable Secur. Comput.*, vol. 9, no. 1, pp. 61–74, Jan. 2012, doi: 10.1109/TDSC.2011.34.

[10] F. Bao and I.-R. Chen, "Dynamic trust management for internet of things applications," in *Proceedings of the 2012 international workshop on Self-aware internet of things - Self-IoT '12*, 2012, p. 1, doi: 10.1145/2378023.2378025.

[11] I.-R. Chen, F. Bao, and J. Guo, "Trust-Based Service Management for Social Internet of Things Systems," *IEEE Trans. Dependable Secur. Comput.*, vol. 13, no. 6, pp. 684–696, Nov. 2016, doi: 10.1109/TDSC.2015.2420552.

[12] F. Bao, I.-R. Chen, and J. Guo, "Scalable, adaptive and survivable trust management for community of interest based Internet of Things systems," in *2013 IEEE Eleventh International Symposium on Autonomous Decentralized Systems (ISADS)*, Mar. 2013, pp. 1–7, doi: 10.1109/ISADS.2013.6513398.

[13] I.-R. Chen, J. Guo, and F. Bao, "Trust Management for SOA-Based IoT and Its Application to Service Composition," *IEEE Trans. Serv. Comput.*, vol. 9, no. 3, pp. 482–495, May 2016, doi: 10.1109/TSC.2014.2365797.

[14] S. K. Prajapati, S. Changder, and A. Sarkar, "Trust Management Model for Cloud Computing Environment," *arXiv.org*, Apr. 2013, [Online]. Available: https://arxiv.org/abs/1304.5313.

[15] C. V. L. Mendoza and J. H. Kleinschmidt, "Mitigating On-Off Attacks in the Internet of Things Using a Distributed Trust Management Scheme," *Int. J. Distrib. Sens. Networks*, vol. 11, no. 11, p. 859731, Nov. 2015, doi: 10.1155/2015/859731.

[16] Y. Ben Saied, A. Olivereau, D. Zeghlache, and M. Laurent, "Trust management system design for the Internet of Things: A context-aware and multi-service approach," *Comput. Secur.*, vol. 39, pp. 351–365, Nov. 2013, doi: 10.1016/j.cose.2013.09.001.

[17] M. Nitti, R. Girau, and L. Atzori, "Trustworthiness Management in the Social Internet of Things," *IEEE Trans. Knowl. Data Eng.*, vol. 26, no. 5, pp. 1253–1266, May 2014, doi: 10.1109/TKDE.2013.105.

[18] J. Renita and N. E. Elizabeth, "Network's server monitoring and analysis using Nagios," in *2017 International Conference on Wireless Communications, Signal Processing and Networking (WiSPNET)*, Mar. 2017, pp. 1904–1909, doi: 10.1109/WiSPNET.2017.8300092.

[19] D. R. E. Lear, R. Droms, "Manufacturer Usage Description Specification, draft-ietf-opsawg-mud-25," 2018. https://tools.ietf.org/html/draft-ietf-opsawg-mud-25 (accessed Apr. 13, 2020).

[20] E. Miehling, M. Rasouli, and D. Teneketzis, "Optimal Defense Policies for Partially Observable Spreading Processes on Bayesian Attack Graphs," in *Proceedings of the Second ACM Workshop on Moving Target Defense - MTD '15*, 2015, pp. 67–76, doi: 10.1145/2808475.2808482.

[21] FIRST, "Common Vulnerability Scoring System version 3.1: Specification Document," 2019. [Online]. Available: https://www.first.org/cvss/specification-document.

[22] E. Hulitt and R. B. Vaughn, "Information system security compliance to FISMA standard: a quantitative measure," *Telecommun. Syst.*, vol. 45, no. 2–3, pp. 139–152, Oct. 2010, doi: 10.1007/s11235-009-9248-8.

[23] United States General Accounting Office, "Information security risk assessment practices of leading organizations," 1998. [Online]. Available: http://www.gao.gov/special.pubs/ai00033.pdf.

[24] I. Kaliszewski and D. Podkopaev, "Simple additive weighting—A metamodel for multiple criteria decision analysis methods," *Expert Syst. Appl.*, vol. 54, pp. 155–161, Jul. 2016, doi: 10.1016/j.eswa.2016.01.042.

[25] M. Modarres and S. Sadi-Nezhad, "Fuzzy Simple Additive Weighting Method by Preference Ratio," *Intell. Autom. Soft Comput.*, vol. 11, no. 4, pp. 235–244, Jan. 2005, doi: 10.1080/10642907.2005.10642907.

[26] N. Griffiths, "Task delegation using experience-based multi-dimensional trust," in *Proceedings of the fourth international joint conference on Autonomous agents and multiagent systems - AAMAS '05*, 2005, p. 489, doi: 10.1145/1082473.1082548.

[27] S. Bhatt, P. K. Manadhata, and L. Zomlot, "The Operational Role of Security Information and Event Management Systems," *IEEE Secur. Priv.*, vol. 12, no. 5, pp. 35–41, Sep. 2014, doi: 10.1109/MSP.2014.103.

[28] H.-J. Liao, C.-H. Richard Lin, Y.-C. Lin, and K.-Y. Tung, "Intrusion detection system: A comprehensive review," *J. Netw. Comput. Appl.*, vol. 36, no. 1, pp. 16–24, Jan. 2013, doi: 10.1016/j.jnca.2012.09.004.

[29] OWASP, "Vulnerability Scanning Tools | OWASP," 2020. https://owasp.org/www-community/Vulnerability_Scanning_Tools (accessed Apr. 27, 2020).

[30] G. Coker *et al.*, "Principles of remote attestation," *Int. J. Inf. Secur.*, vol. 10, no. 2, pp. 63–81, Jun. 2011, doi: 10.1007/s10207-011-0124-7.

[31] NIST, "National Vulnerability Database (NVD)," 2020. https://nvd.nist.gov/ (accessed Jun. 24, 2020).

[32] C. Vassilakis, C.-M. Mathas, and N. Kolokotronis, "Cyber-Trust Project D5.4: Trust management service: security and privacy," 2020. [Online]. Available: https://cyber-trust.eu/project-deliverables/.